# Broadband second harmonic generation in whispering gallery mode resonators

Guoping Lin[1], Josef U. Fürst[1,2], Dmitry V. Strekalov[1], and Nan Yu[1,*]

[1]*Jet Propulsion Laboratory, California Institute of Technology, Pasadena, California 91109, USA*
[2]*Institut für Optik, Information und Photonik, University of Erlangen-Nuremberg, Erlangen 91058, Germany*
[*] *nan.yu@jpl.nasa.gov*

*Optical frequency conversion processes in nonlinear materials are limited in wavelength by the accessible phase matching and the required high pump powers[1-4]. In this letter, we report a novel broadband phase matching (PM) technique in high quality factor (Q) whispering gallery mode (WGM) resonators[5-8] made of birefringent crystalline materials. This technique relies on two interacting WGMs, one with constant and the other with spatially oscillating phase velocity. Thus, phase matching occurs cyclically. The technique can be implemented with a WGM resonator with its disk plane parallel to the optic axis of the crystal. With a single beta barium borate (BBO) resonator in that configuration, we experimentally demonstrated efficient second harmonic generation (SHG) to harmonic wavelengths from 780 nm in the near infrared to 317 nm in the ultraviolet (UV). The observed SHG conversion efficiency is as high as 4.6% (mW)-1. This broadband PM technique opens a new way for nonlinear optics applications in WGM resonators. This work is also the first reported continuous wave UV generation by direct SHG in a WGM resonator.*

Frequency conversion through nonlinear optical processes has been ubiquitously used to generate laser light at spectral regions from UV to far infrared where direct laser light is not available. Ever since the first SHG by Franken et al. in 1961, it is recognized that an efficient frequency conversion depends critically on both phase matching (PM) and pump light intensity. Until now, birefringent PM and quasi-phase matching (QPM) are two main techniques successfully used[1-4]. The former relies on incidental PM between the pump and the harmonics having different polarizations. The latter typically requires artificial patterning of the optical nonlinearity of the material. The most commonly used patterning method is periodic poling which unfortunately excludes most nonlinear crystals. Recent research efforts have explored various PM conditions such as modal dispersion[9-11] and form birefringence[12,13].

The high pump intensities needed for efficient conversion are typically achieved through temporally confined laser pulses, spatially confined light in waveguides, or continuous wave (cw) laser power enhancement in resonant cavities[14-17]. The latter two strategies can be employed in WGM resonators[5-8] as WGM resonators are capable of supporting high-Q modes and small mode volumes by continuous total internal reflections in a spheroidal dielectric material. Modal dispersion PM[18], non-critical PM[19] and periodically poled QPM have been explored in nonlinear WGM resonators typically made from LiNbO$_3$[20-23]. Periodically patterning the surface with molecular layers using QPM on WGM resonators was also demonstrated[24]. Likewise, all PM methods have very limited wavelength bandwidth and the ferroelectric crystal materials are not suitable for UV applications.

Motivated by exploring new PM schemes and diverse nonlinear UV crystals, we investigated WGM resonators made of uniaxial BBO crystals[25] with their optic axis tilted instead of parallel to the resonator symmetry axis[26] (often referred to as a z-cut resonator[19-23]). A non z-cut resonator geometry results in a significant variation of the refractive index along the disk's circumference which allows PM. This is akin to reported QPM methods in AlGaAs and GaAs micro resonators[27,28]. However, we have found that only resonator disks with the optic axis parallel to the plane of the disk (thereafter "xy-cut") support two orthogonally polarized modes. This is the resonator geometry being considered here.

The two orthogonally polarized modes can in general be characterized as TE modes with their polarization perpendicular to the disk plane and TM modes parallel to it as shown in Fig. 1a. The xy-cut breaks the typical rotational symmetry of a WGM resonator. The polarization of TE modes remains orthogonal to the optic axis as light travels; its refractive index is constant. In contrast, the refractive index of TM modes is expected to vary as the polarization rotates relative to the optic axis along the circumference of the disk. The refractive index of the TM modes can be calculated with the expression:

$$\frac{1}{n^2(\lambda,\theta)} = \frac{cos^2\theta}{n_o(\lambda)^2} + \frac{sin^2\theta}{n_e(\lambda)^2} \qquad (1)$$

where $n_o(\lambda)$ and $n_e(\lambda)$ are refractive indices of ordinary and extraordinary polarized rays respectively available from Sellmeier equations[29], $\theta$ is the angle between the wave vector and the optic axis. The indices evaluated for 1557 nm and its second harmonic are plotted in Fig. 1b. The refractive index of TM rays oscillates from the ordinary index value $n_o$ to the extraordinary index value $n_e$ while the refractive index of TE rays is a constant as expected. More significantly, while the phase mismatch between the TE pump field and TM second harmonic field oscillates, there are crossover points in each round trip where the phase between the TE fundamental and TM second harmonics are matched in the xy-cut BBO resonator.

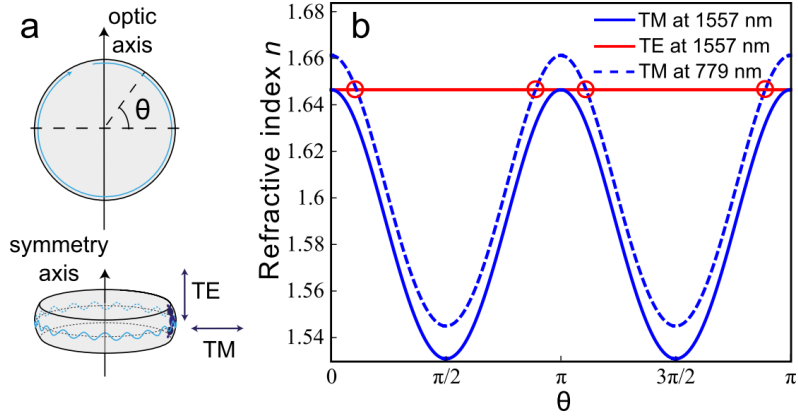

**Figure 1** Refractive indices of WGMs in an xy-cut BBO disk resonator. a, Illustration of the optic axis z, the relative angle q, and TE and TM polarizations in a disk resonator. b, Plot of the refractive indices of WGMs along the disk circumference for TE and TM at 1557 nm pump fundamental and its TM second harmonic. There are four regions (red-circled) where phase matching occurs between the TE fundamental field at 1557 nm and its TM second harmonic at 779 nm.

The generation of second harmonic field is described in general by the equation:

$$\frac{dE_2}{dz} = \frac{2i\omega_2^2}{k_2 c^2} d_{eff} E_1^2 \, e^{i \int_0^z \Delta k dz} \qquad (2)$$

where $\Delta k = 2k_1 - k_2$ represents the phase difference between the fundamental wave vector $k_1 = 2\pi n_1/\lambda_1$ and that of the second harmonic field $k_2 = 2\pi n_2/\lambda_2$, with their corresponding wavelengths $\lambda_1 = 2\lambda_2$ and refractive indices $n_1$ and $n_2$. $\omega$, c and $d_{eff}$ represent the corresponding angular frequency, the speed of light, and the effective nonlinear coefficient respectively. The condition $\Delta k = 0$ represents perfect PM. In this case, the second harmonic field builds up constructively as the pump and the second harmonic copropagate along the nonlinear material. This ideal PM condition is illustrated in Fig. 2a. Alternatively when such an ideal condition cannot be met in a given crystal, one can artificially reverse the sign of the effective nonlinear coefficient periodically to accomplish QPM in ferroelectric materials, as illustrated in Fig. 2a. In this case, the second harmonic field can grow, though not as quickly as in the perfect phase matching case.

In the xy-cut WGM geometry both phase mismatch $\Delta k$ and the effective nonlinear coefficient $d_{eff}$ oscillate as a function of $\theta$. The $\theta$-dependent oscillation in general results in four phase matched regions per round trip and an overall favorable phase matching condition. Additionally, the doubly resonant condition (for the pump and for the second harmonic) ensures the repeated phase matching and the growth of the second harmonic field in the high Q resonator. We call this cyclic semi-phase matching (CSPM) condition, referring the PM regions in parts of each round trip and the recurring process with circulating waves in the xy-cut resonator. This picture is also illustrated in Fig. 2a.

To understand the phase matching process more quantitatively, we need to carry out the integration of Equation (2). We will make a further simplification by unfolding the circulating path in a resonator to a linear waveguide of a length equal to the effective path length of the resonator. This approximation works particularly well when considering the doubly resonant condition where the light field phases of both pump and second harmonic repeat exactly after one round trip. One thus needs only to evaluate the field gain of a single round trip to understand the PM for the harmonic conversion. This single path length is about 7 mm for the xy-cut crystal resonator used in our study. By numerically integrating equation (2) with an undepleted pump, we evaluated the relative conversion efficiencies for two fundamental wavelengths around 1557 nm and 634 nm. We found indeed that this model predicts phase matching over a wide pump wavelength range, with the short wavelength limit of 410 nm determined by the birefringence of BBO and the long wavelength limit of 2.6 μm due to the material absorption. The conversion efficiency however is modulated, so the optimal phase matching is achieved only at discreet wavelength regions. The optimal conversion frequencies can be tuned simply by changing the optical path length around the resonator. By changing the dimension of the resonator by a mere 0.2%, one can shift a minimum conversion point to a maximum, as shown in Fig. 2b. Alternatively, optimal wavelengths can be found by coupling different WGMs exhibiting different effective resonator lengths. This shows that CSPM is indeed applicable in a broad wavelength range.

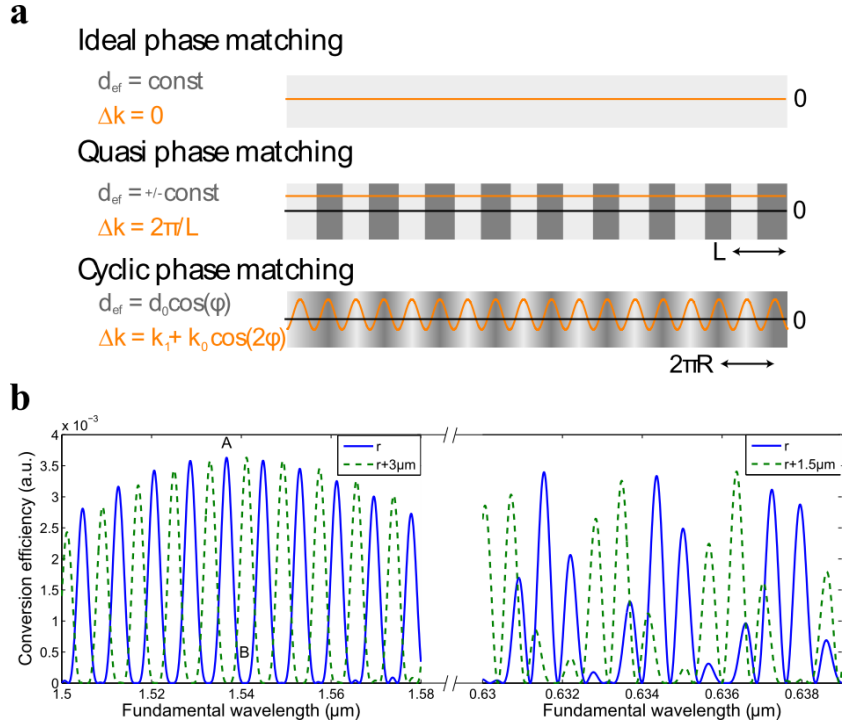

**Figure 2.** Illustration of phase matching schemes including the cyclic semi-phase matching. a, Illustration of various phase matching schemes in terms of the nonlinear coefficients $d_{eff}$ and the phase mismatch $\Delta k$. Light propagates in the horizontal direction. In the ideal phase matching condition, $d_{eff}$ is a constant while $\Delta k$ is uniformly zero, i.e. phase matched; in quasi-phase matching, the sign of $d_{eff}$ is altered with a periodicity such that the second harmonic gain is guaranteed; In cyclic semi-phase matching, both $d_{eff}$ and $\Delta k$ vary and show discreet phase matched regions along the circumference. b, Calculated conversion efficiencies for CSPM in a single round trip for two wavelength regimes as a function of pump wavelength. The dashed green curves shown the tuning of the optimal conversion regions by changing the resonator radius $r$.

It is interesting to study the phase evolution and the growth dynamics of the second harmonic field in the xy-cut resonator. We have plotted separately the calculated phase mismatch and the second harmonic amplitude as a function of the angle $\theta$ and the corresponding amplitude diagram in the complex plane in Fig. 3. The color variation of the amplitude plots represents the propagation along the angle $\theta$. Fig. 3 a) and 3 b) correspond to maximal and minimal conversion efficiencies as indicated in Fig. 2b. One can see that the second harmonic field strength oscillates rapidly with small amplitudes when the phase mismatch is large while it increases or decreases effectively when the phase mismatch disappears. In the complex plane, the second harmonic amplitude growth is trapped in a smaller spiral for a larger phase mismatch but spreads out quickly as the phase mismatch disappears. In general we have four short effective nonlinear coupling regions per round trip. Their locations determine either constructive or destructive interferences for second harmonic generation in these four regions. The separations among these four regions have two characteristic length scales (see Fig. 1) which result in a convolution of a slow and a fast modulation in the conversion efficiency for different wavelengths. In Fig. 2b, this is most prominent for shorter wavelengths.

Phase matching in the short discrete regions is less efficient compared to the ideal and quasi phase matching. In a high Q resonator, this deficit is partly made up by the long effective path length and the coherent double resonance condition. In addition, the CSPM scheme uses natural phase matching characteristics of the host material and does not require any additional material processing such as periodical poling. QPM may even be combined with CSPM to make CSPM an even more flexible tool.

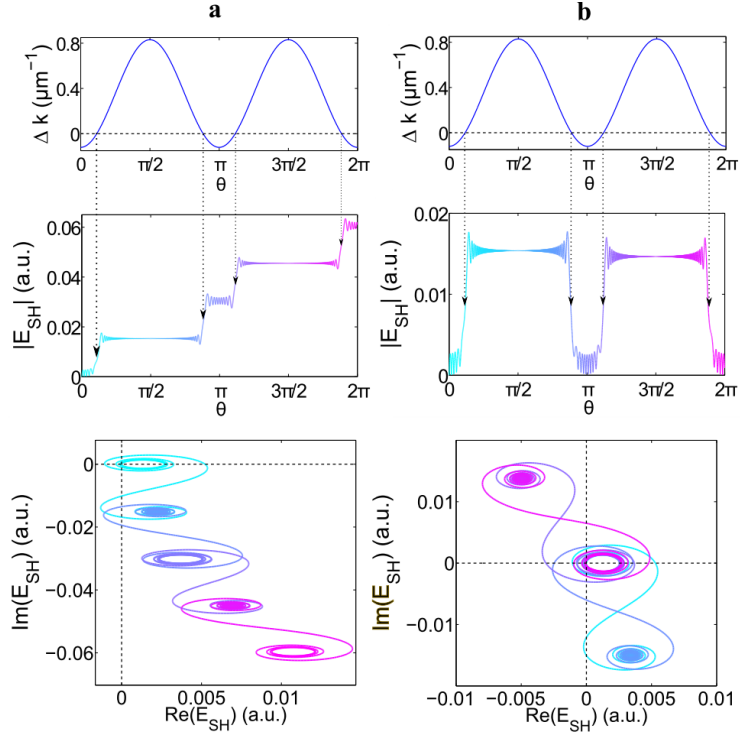

**Figure 3** | Growth and dynamics of the second harmonic amplitude for cyclic semi-phase matching. From top to bottom: calculated phase mismatch $\Delta k$ versus angle $\theta$ around the disk resonator, electric field amplitude of the second harmonic signal vs. $\theta$, and the second harmonic growth in the complex plane (phasor plot) where the color of the curves indicates the angular position $\theta$. The left and right panels a and b show a maximum and a minimum conversion case respectively.

For experimental verification of the CSPM for SHG, we fabricated an xy-cut BBO WGM resonator from a crystalline disk substrate with its edge polished into a spheroid shape of optical grade smoothness[8]. The disk has a diameter of about 1.82 mm and a few hundred microns thickness. Though the optic axis induced unequal surface stiffness along the circumference, we were able to produce disks with measured Q factors on the order of $10^7$ from near infrared (1560 nm) up to UV (370 nm). The resonator is mounted in a setup illustrated in Fig. 4a. A tunable laser source at 634 nm was first used as the fundamental pump source. A linear polarizer and a half-wave plate are used to control the input laser polarization. The TE polarized pump beam is then focused into a sapphire prism for evanescent excitation of WGMs in the BBO disk. The output beam from the prism is collimated with a second lens and sent through a dichroic mirror to separate the fundamental and second harmonic signals which are monitored simultaneously with detectors. A bandpass filter is placed in front of the second harmonic detector to further remove noise from the pump signal.

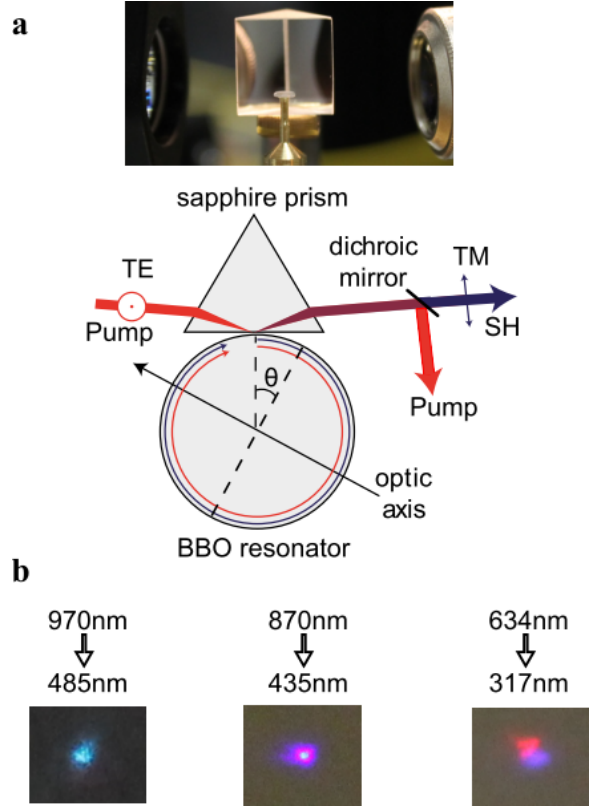

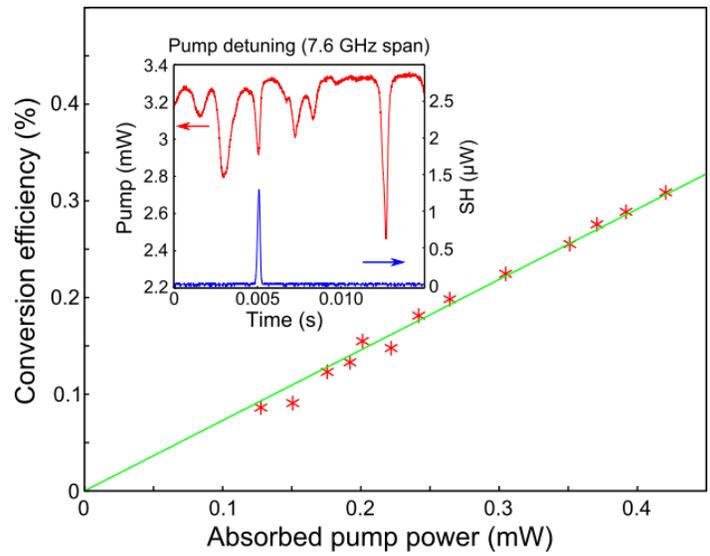

**Figure 4** | Experimental setup for the second harmonic generation. a, Schematic of the experimental setup with a photo of the resonator-prism coupling setup. The BBO disk resonator mounted on a brass post is at the front of the comparatively coupling prism. b, True color images of the second harmonic beam profiles generated from different pump wavelengths.

**Figure 5** | Pump power dependence of the second harmonic conversion efficiency measured at 634 nm. The linear dependence confirms the second order nonlinear process for the second harmonic generation. Inset: the transmitted pump signal and the second harmonic signal as the pump laser frequency scanned across a second harmonic generation modes. The linewidth of the second harmonic appears narrower than the pump because of the nonlinear pump power dependence of the second harmonic.

The WGM resonator is pumped with a few milliwatts of power. The pump laser frequency is scanned over several WGMs and the output signals are detected simultaneously. The inset of Fig. 5 shows an example of spectra obtained when scanning the pump laser frequency across several TE WGMs at 634nm. The second harmonic signal appears when the pump laser is swept over one of the pump modes where the double resonance condition and the CSPM condition are fulfilled. Stronger second harmonic signals are usually obtained when the WGM at pump is over-coupled, as the SH signal couples out more efficiently for smaller gap distances[30]. The SHG has been further confirmed by the linear pump power dependence of the second harmonic conversion efficiency as shown in Fig. 5. To obtain this set of data, the pump was over-coupled. The in-coupled fundamental pump power is estimated by fitting the dip and obtaining the height of the resonance dip in the pump transmission. We varied the incident pump power and kept the coupling gap unchanged during this measurement.

We also used cw lasers at 870 nm, 974 nm and 1557 nm to pump the same BBO resonator. We observed high conversion efficiencies in all these cases. Fig. 4b shows the images of the second harmonic output beams observed on a paper card inserted behind the dichroic mirror or band pass filter. The violet and blue-green correspond to the second harmonic signal at 435 nm and 487 nm. The blue and red spots are related to luminescence from 317 nm on the paper card and the residual pump laser at 634 nm respectively. The spatial separation between blue and red is due to the dispersions in the WGMs in the resonator and the coupling prism. For clarity, we summarize the best conversion efficiencies obtained in the 1.82 mm diameter BBO resonator for all the wavelengths used in Table 1. Compared with previously reported cavity-enhanced SHG systems at UV wavelengths[16,17], the demonstrated CSPM efficiencies are comparable and higher. If one compares to a comparably sized setup and broadbandness achievable from a single pass crystal, however, this improvement factor is several orders of magnitude.

It should be noted that the observed efficiencies are affected by three main factors related to the WGM resonator: the Q factors of the pump and the harmonic modes, the spatial and spectral overlap factors between these two WGMs, and the overall CSPM condition. We have estimated the expected conversion efficiency for the experimental BBO setup using the method described in References (19) and (20). In addition to those parameters tabulated in Table 1, we assumed a loaded Q factor of $5\times10^6$ at second harmonic wavelengths and an average nonlinearity of $d_{22}(0.003)^{1/2}$ where $d_{22}$ is the maximum nonlinearity tensor of BBO crystal that can only be achieved in the case of x-cut resonator geometry. The factor $(0.003)^{1/2}$ comes from Fig. 2b that is normalized to a complete phase matching condition. The efficiency is also affected by the spatial mode overlap between the fundamental and second harmonics, thus we used a similar mode overlap factor of 0.3 as in Reference (20). The results are also shown in Table 1. For the highest achieved efficiency of 4.6%/mW at 974 nm, it is within a factor of 20 from the estimated value. Given the high sensitivity to the $Q$s, unknowns of the mode overlap and the probably over-estimated effective nonlinearity from $d_{22}$, the discrepancy is not unreasonable. Large variations at different wavelengths are likely due to the position on Fig.2 (b). Nevertheless, phase matching must exist in the resonator even for the 0.063%/mW conversion efficiency, the lowest observed for all the wavelengths studied.

The observed SHG over more than one octave in wavelength in a single xy-cut crystal is consistent with the prediction of the CSPM process. One of the key phenomena making the CSPM possible is the existence of the TM mode with varying refractive index. We experimentally confirmed that the polarization of the outgoing second harmonic beam is along the expected TM direction, in the plane of the disk resonator. In fact, by rotating the disk and therefore the relative coupling point with respect to the optic axis, we observed the angular periodic variation of the SHG output beam spot, as expected from the $\theta$ dependence of the refractive index of a TM WGM[31]. In addition the measured free spectral range of the TM polarized WGMs yields an average refractive index of 1.59 (±0.01) that matches theoretical value of 1.587 for TM WGMs as shown in Fig. 1b. On the other hand, when we pumped with the same light but in the TM polarization direction in the same resonator, we did not observe any second harmonic signal. This is consistent with our model, as there exist no phase matching regions between a TM pump mode and a TE second harmonic. This strongly supports the theory of CSPM. To further demonstrate the critical role of the phase matching points, we investigated both a WGM resonator from z-cut BBO and one from x-cut quartz. In both cases, there are no phase matching regions found between TE pump and TM second harmonic. In both cases, we were unable to observe any second harmonic signals under similar conditions.

**Table 1. SHG conversion efficiency for different pump wavelengths**

| Fundamental/ second harmonic Wavelengths (nm) | Pump (mW) | Second harmonic power (μW) | Loaded Q factors at the fundamental wavelength | Conversion efficiency $P_{SH}/P_{in}^2$ (mW$^{-1}$) |
|---|---|---|---|---|
| 1557 / 778 | 1.1 | 0.76 | $3\times10^6$ | 0.063% |
| 974 / 487 | 0.25 | 2.9 | $7\times10^6$ | 4.6% |
| 870 / 435 | 0.87 | 13.5 | $1\times10^7$ | 1.8% |
| 634 / 317 | 0.42 | 1.3 | $4\times10^6$ | 0.74% |

**\* After the same BBO resonator was repolished.**

In summary, we have theoretically developed and experimentally confirmed the broadband technique of cyclic semi-phase matching in a crystalline WGM resonator with its optic axis parallel to the resonator plane. We demonstrated for the first time the remarkable phase matching properties in this configuration and its applications to nonlinear optics. With a single resonator of xy-cut BBO, we were able to show efficient SHG over more than one octave in wavelength from the near infrared, through the visible up to the ultraviolet wavelengths regime. In theory, cyclic semi-phase matching will cover the entire birefringent phase matching range of the nonlinear crystal used. Choosing BBO as the host material here, we have also made the first observation of cw SHG in the UV regime in WGM resonators. The cyclic semi-phase matching paves the way for a broader array of nonlinear optics applications with WGM resonators for both cw and mode-locked lasers. Using new nonlinear crystals such as KBBF[32], it is now possible to extend the SHG into the vacuum UV wavelength regime.

**Acknowledgments**
This work was performed at the Jet Propulsion Laboratory, California Institute of Technology, under a contract with NASA. G. Lin acknowledges support from the NASA Postdoctoral Program, administered by Oak Ridge Associated Universities (ORAU). J. U. Fürst acknowledges financial support from the Max Planck Society and thanks Rosalin Hertrich for valuable comments.